\documentclass[preprint,
superscriptaddress,
 amsmath,amssymb,
 aps,
]{revtex4-2}

\usepackage{amssymb}
\usepackage{epsfig}
\usepackage{esint}
\usepackage{xcolor}

\begin{document}

\pagenumbering{gobble}

\title{Contact lines on stretched soft solids: Modeling anisotropic surface stresses}

\author{Stefanie Heyden}
 \email{stefanie.heyden@mat.ethz.ch}
 \affiliation{Department of Materials, ETH Z\"{u}rich, 8093 Z\"{u}rich, Switzerland.}
 
 \author{Nicolas Bain}
 \affiliation{Department of Materials, ETH Z\"{u}rich, 8093 Z\"{u}rich, Switzerland.}
 
  \author{Qin Xu}
 \affiliation{Department of Physics, The Hong Kong University of Science and Technology, Clear Water Bay, Kowloon, Hong Kong.}

\author{Robert W. Style}
 \affiliation{Department of Materials, ETH Z\"{u}rich, 8093 Z\"{u}rich, Switzerland.}
 
\author{Eric R. Dufresne}
 \email{eric.dufresne@mat.ethz.ch}
 \affiliation{Department of Materials, ETH Z\"{u}rich, 8093 Z\"{u}rich, Switzerland.}

\begin{abstract}
We present fully analytical solutions for the deformation of a stretched soft substrate due to the static wetting of a large liquid droplet, and compare our solutions to recently published experiments (Xu \emph{et al}, \emph{Soft Matter} 2018). 
Following a Green's function approach, we extend the surface-stress regularized Flamant-Cerruti problem to account for uniaxial pre-strains of the substrate. 
Surface profiles, including the heights and opening angles of wetting ridges, are provided for linearized and finite kinematics.
We fit experimental wetting ridge shapes as a function of applied strain using two free parameters, the  surface Lam\'{e} coefficients.
In comparison with experiments, we find that observed opening angles are more accurately captured using finite kinematics, especially with increasing levels of applied pre-strain. 
These fits qualitatively agree with the results of Xu \emph{et al}, but revise values of the surface elastic constants.

\end{abstract}

\maketitle

\section{Introduction}

Capillary phenomena of liquids are familiar to us from everyday life and have a long history in theoretical mechanics \cite{Maxwell:1878}.  
There, surface stresses dominate the mechanics of liquids, leading to a range of phenomena from the spherical shape of small water droplets to capillary action within narrow tubes.
Their effects are visible to the naked eye because their characteristic scale is set by the capillary length $L_c=\sqrt{\gamma/\rho g}\sim\mathcal{O}(\mathrm{mm})$, which relates the surface energy, $\gamma$,  to density, $\rho$, and gravitational acceleration, $g$.
In solids, capillary phenomena dominate at length scales much shorter than the elastocapillary length, $l_e\sim\Upsilon/E$.
Here, $\Upsilon$ is the surface stress and $E$ denotes the material's Young's modulus. 
For stiff engineering materials ($E\sim\mathcal{O}(\mathrm{GPa})$), the elastocapillary length is comparable to the atomic scale. 
Thus, capillary phenomena are rarely significant in stiff materials. 
Soft materials ($E\sim\mathcal{O}(\mathrm{kPa})$) feature elastocapillary lengths that can  reach $\mathcal{O}(10~\mathrm{\mu m})$.
Thus, capillary phenomena can become relevant  in a variety of settings, \emph{e.g.}, the wetting behavior on soft substrates \citep{Carre:1996,Extrand:1996,Gennes:2004,Roman:2010,Yu:2012}, flattening of nearly plane soft solids \citep{Jagota:2012}, stiffening effects in composites with liquid inclusions \citep{Style:2015}, or soft adhesion \citep{Style:2015}. 

Solid surface stresses have several features that distinguish them from liquid surface stresses.
Due to the ability of liquid molecules to freely rearrange, the surface energy of simple liquids is independent of deformation.
Consequently, the surface stress takes on a strain-independent isotropic form, $\Upsilon_{ij} = \gamma^l~ \delta_{ij}$.
Thus, the terms `surface tension' and `surface energy' are used interchangeably in that context  \citep{Style:2017}. 
Solids on the other hand exhibit much more complex surface stresses, including, \emph{e.g.}, anisotropy \citep{Stein:2018,Dreher:2019} and inelasticity \citep{Altenbach:2012,Pepicelli:2019}. 
Our understanding of the underlying physical principles governing surface stresses in soft solids, however, is only weakly developed and constitutes an active subject of research \citep{Jensen:2017,Xu:2017,Xu:2018,Schulman:2018,Masurel:2019, Daniels:2014,Cao:2014,Andreotti:2020}.

\begin{figure}[t]
\begin{center}
    \includegraphics[width=0.9\linewidth]{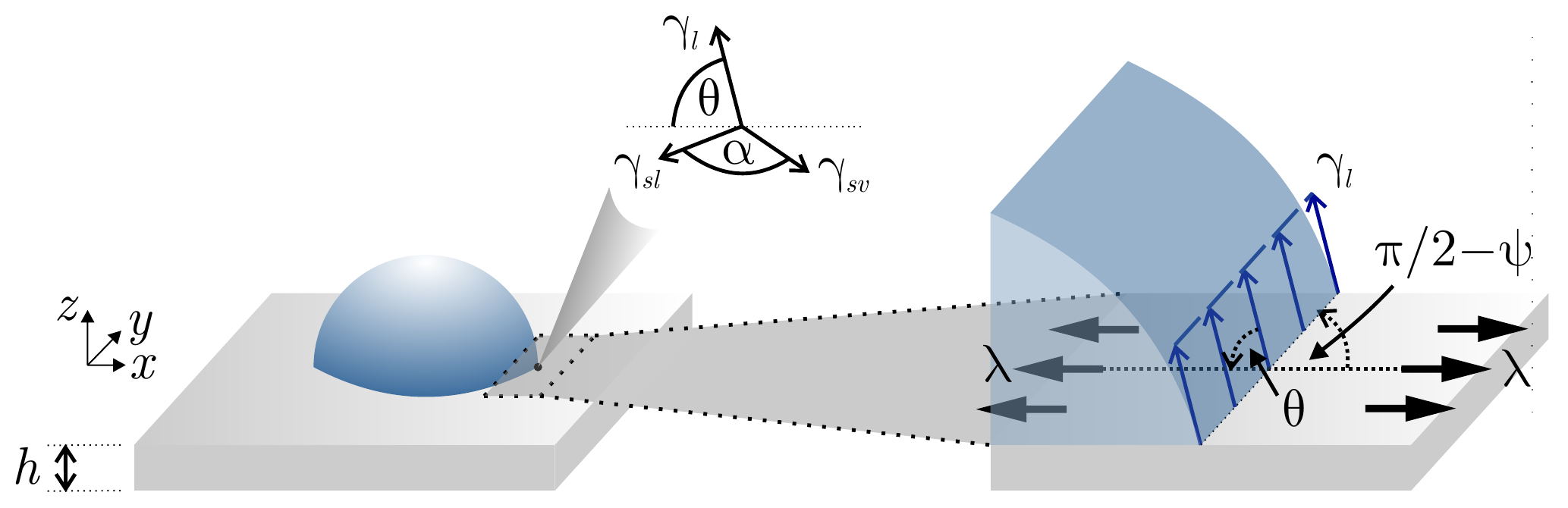}
    \caption{Sketch of the static wetting problem under consideration, illustrating opening angle $\alpha$ of the wetting ridge, contact angle $\theta$, as well as in-plane angle $\psi$ between an applied pre-strain $\epsilon_{xx}$ and the contact line generated by liquid surface tension $\gamma_l$. 
    }
\label{Fig:Droplet}
\end{center}
\end{figure}

To shed light on surface stresses in soft solids, recent experiments have used static wetting techniques  \citep{Style:2013, Style:2017, Xu:2017, Xu:2018}. 
There, liquid droplets were deposited on  soft substrates, and the microscopic geometry of the contact-line (where the three interfaces meet)  was precisely measured.
As illustrated on the left of Figure \ref{Fig:Droplet}, surface stresses were inferred  from a local force balance, ignoring any contributions from bulk elasticity. The net force per unit length of the contact line was assumed to have contributions from the surface tensions of the liquid-vapor interface,  $\gamma^l\hat{t}_i^{sl}$, the solid-liquid interface,  $\Upsilon_{ij}^{sl}\hat{t}_j^{sl}$,  and the solid-vapor interface, $\Upsilon_{ij}^{sv}\hat{t}_j^{sv}$. Here, $\Upsilon_{ij}$ is the surface stress tensor and $\hat{t}_j$ is the tangent vector of each interface. To gain insights into the strain-dependence of surface stresses, recent experiments \citep{Xu:2017,Xu:2018} have applied the same approach to stretched substrates as shown in Figure~\ref{Fig:Droplet} on the right. In these experiments, the substrate is stretched in both uniaxial \citep{Xu:2017} as well as biaxial \cite{Xu:2018} set-ups preceding droplet deposition.

\begin{figure}[t]
        \begin{minipage}{.5\textwidth}
        \centering
        \includegraphics[width=0.99\textwidth]{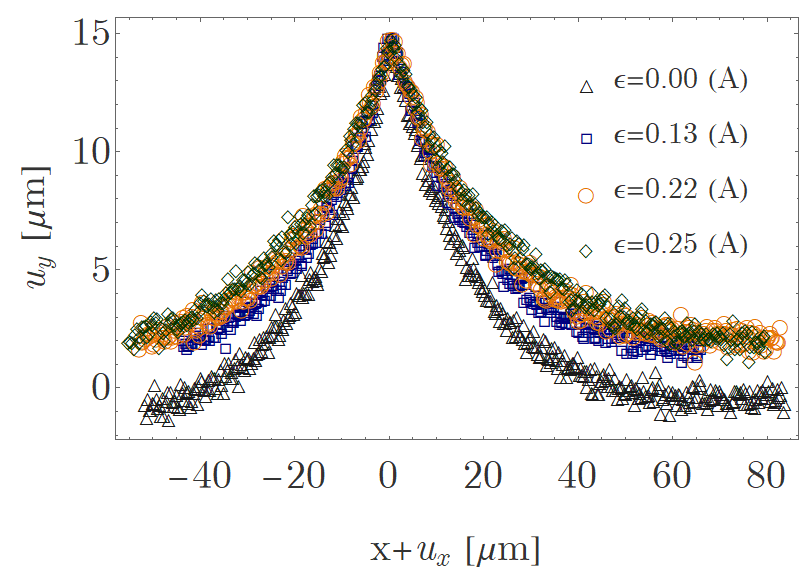}
    \end{minipage}%
    \begin{minipage}{0.5\textwidth}
        \centering
        \includegraphics[width=0.99\textwidth]{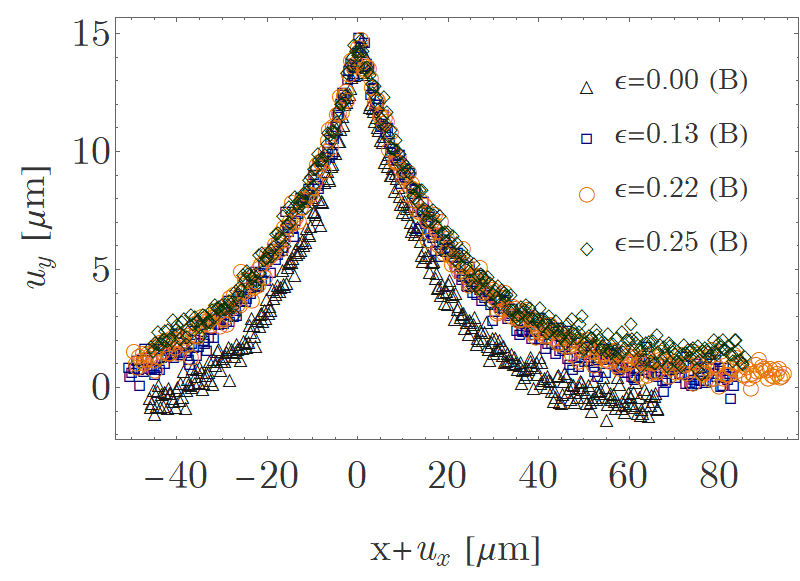}
    \end{minipage}
    \begin{minipage}{.5\textwidth}
        \centering
        \includegraphics[width=0.99\textwidth]{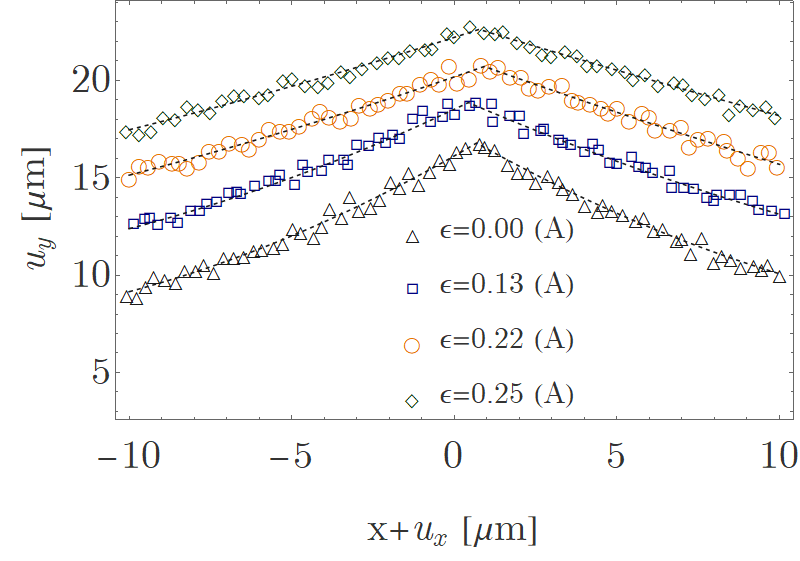}
    \end{minipage}%
    \begin{minipage}{0.5\textwidth}
        \centering
        \includegraphics[width=0.99\textwidth]{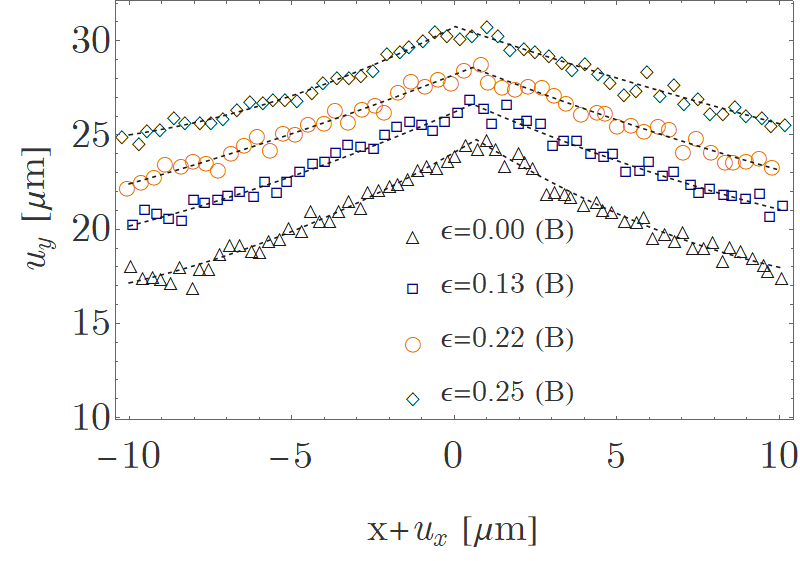}
    \end{minipage}
\caption{Top: Experimental static wetting profiles within the direction of applied pre-strain (A) and perpendicular to the direction of applied pre-strain (B). Bottom: Zoom into the surface stress dominated region of experimental static wetting profiles and corresponding fittings. The data shown here is a subset of experimental static wetting profiles first published in \citep{Xu:2018}}.
\label{Fig:Profiles}
\end{figure}

Data from \cite{Xu:2018} is shown in Figure \ref{Fig:Profiles}, which depicts the measured static wetting profiles within the direction of applied pre-strain and perpendicular to the direction of applied pre-strain. The slope of the surface converges to a finite value below the elastocapillary length since the tip behavior of wetting ridges is dominated by surface stresses \citep{Wu:2018}.
Thus, the tip geometry is well-described by the opening angle, $\alpha$.
As shown in \citep{Xu:2018}, the opening angles are strain-dependent.
A balance of surface stresses at the tip suggests that the surface stresses are also strain dependent. 
An alternate explanation of these data has emerged recently, based on nonlinearities within the bulk material, propagated to singular points at the surface  \cite{Masurel:2019}. 

Analytical solutions to the static wetting problem trace back to classical problems of point- and line forces on an elastic half space commonly referred to as the Boussinesq- and Flamant problem, respectively \citep{Boussinesque:1885,Flamant:1892}. In addition, Cerruti's problem considers a point force that is tangent to the surface of a half-space \citep{Cerruti:1882}. Solutions were subsequently extended to account for anisotropic half-spaces \citep{Sveklo:1964,Willis:1967}, as well as isotropic media of finite thickness and possibly multiple layers \citep{Burmister:1945,Yue:1996,Merkel:2007}. 

Here, we extend the analytical solution to the static wetting problem in the large droplet limit to furthermore account for pre-strains of the surface. To predict the fully three-dimensional profile of the wetting ridge, solutions account for varying in-plane angles between the attendant line force and the applied pre-strain. To gain further insight into the characteristics of the strain-dependence of surface stresses, the theory is laid out in both frameworks of linearized and finite kinematics and predicted opening angles of the wetting ridge are compared to available experimental data.
We compare our results to published experiments, finding good agreement for data with revised elastic constants.

\section{Formulation of the problem}

To model substrate deformations upon static wetting in the limit of large droplets, we consider a substrate of finite thickness, $h$, as shown in Figure~\ref{Fig:Droplet}. 
The substrate is assumed to be neutrally-wetted, such that $\Upsilon_{ij}^{sv}=\Upsilon_{ij}^{sl}$. For convenience, both will be referred to as $\Upsilon_{ij}$ in the following. 
We further allow for an emergent anisotropy in surface stresses due to pre-strains applied to the substrate. Solutions account for varying in-plane angles $\psi$ between the applied pre-strain and the in-plane direction perpendicular to the contact line, thus predicting the fully three-dimensional surface profile of a large sessile droplet.

\subsection{Bulk elasticity}

We follow a Green's function approach in studying the effect of a line force acting upon an elastic substrate within the linearized kinematics framework. Starting from linear momentum balance, insertion of the Cauchy relations for an isotropic linear elastic material gives the well-known Navier-Lam\'{e} equations
\begin{align}
	(\lambda + 2\mu)u_{x,xx} + (\lambda+\mu)(u_{y,yx}+u_{z,xz}) + \mu(u_{x,yy} + u_{x,zz}) &= 0 \nonumber \\
	(\lambda + 2\mu)u_{y,yy} + (\lambda+\mu)(u_{x,xy}+u_{z,yz}) + \mu(u_{y,xx} + u_{y,zz}) &= 0 \label{eq:Navier}  \\
	(\lambda + 2\mu)u_{z,zz} + (\lambda+\mu)(u_{x,xz}+u_{y,zy}) + \mu(u_{z,xx} + u_{z,yy}) &= 0, \nonumber
\end{align}
whereby gravitational effects are neglected. For the limiting case of large droplets, the line force generated by the liquid surface tension $\gamma_l$ may be approximated as a straight contact line as illustrated in Figure~\ref{Fig:Droplet}. Under the assumption of plane strain within the in-plane direction parallel to the contact line, the set of Equations~\eqref{eq:Navier} simplifies to
\begin{align}
	(\lambda + 2\mu)u_{x,xx} + (\lambda + \mu)u_{z,xz} + \mu u_{x,zz} &= 0 \nonumber \\
	(\lambda + 2\mu)u_{z,zz} + (\lambda + \mu)u_{x,xz} + \mu u_{z,xx} &= 0.
\label{eq:Navier_Simplified}
\end{align}
Application of Fourier transformations in $x$ leads to a system of ordinary differential equations, which can be recast into matrix form as
\begin{equation*}
\begin{pmatrix}
    \hat{u}_{x,z} \\
	\hat{u}_{z,z} \\
	\hat{u}_{x,zz} \\
	\hat{u}_{z,zz} 
  \end{pmatrix} = \begin{pmatrix}
	0 & 0 & 1 & 0 \\
	0 & 0 & 0 & 1 \\
	\frac{(\lambda+2\mu)}{\mu}k^2 & 0 & 0 & -\frac{i(\lambda+\mu)}{\mu}k \\
	0 & \frac{\mu}{(\lambda+2\mu)}k^2 &-\frac{i(\lambda+\mu)}{(\lambda+2\mu)}k & 0
\end{pmatrix}\begin{pmatrix}
    \hat{u}_{x} \\
	\hat{u}_{z} \\
	\hat{u}_{x,z} \\
	\hat{u}_{z,z} 
  \end{pmatrix}
\end{equation*}
with wave number $k$. In shorthand notation, a solution to the system 
\begin{equation}
\frac{\partial}{\partial z}\hat{u}_j = A_{ij}\hat{u}_j
\end{equation}
may be written in terms of matrix exponentials 
\begin{equation}
	\hat{u}_j = e^{A_{ij}z}\hat{u}^0_j,
\end{equation}
whereby we find the Jordan normal form of $A_{ij}$ to compute the matrix exponential $B_{ij}=e^{A_{ij}z}$ (the full form of $B_{ij}$ is given in the appendix). Here, $\hat{u}^0_j$ denotes boundary conditions applied at the bottom of the substrate in the form $\hat{u}_x^0=0$ and $\hat{u}_z^0=0$. The solution may be reformulated as
\begin{equation}
	\hat{u}_i = B_{ij}\hat{u}^0_j = \begin{pmatrix}
    B_1 & B_2 \\
    B_3 & B_4 
  \end{pmatrix} \hat{u}^0_j,
\label{eq:Bij}
\end{equation}
with $\hat{u}_i = (\hat{u}_x,\hat{u}_z,\hat{u}_{x,z},\hat{u}_{z,z})$ and $\hat{u}^0_i = (0,0,\hat{u}^0_{x,z},\hat{u}^0_{z,z})$. Each block $B_i$ furthermore constitutes a $2\times2$ matrix, with 
\begin{equation}
   \begin{pmatrix}
    \hat{u}_x \\
    \hat{u}_z
  \end{pmatrix} = B_2 \begin{pmatrix}
    \hat{u}_{x,z}^0 \\
    \hat{u}_{z,z}^0.
  \end{pmatrix}
\end{equation}
Using this relation, we can replace $\hat{u}^0_j$ with $(0,0,B_2^{-1}(\hat{u}_x,\hat{u}_z))$, which gives
\begin{equation}
	\hat{u}_i = \begin{pmatrix}
    B_2\,B_2^{-1} \\
    B_4\,B_2^{-1}
  \end{pmatrix} \begin{pmatrix}
    \hat{u}_x \\
    \hat{u}_z 
  \end{pmatrix}.\end{equation}

In the following, we aim to solve for substrate displacements at the surface $z=h$ only. Note that in the absence of uniaxial pre-strains $\epsilon_{xx}^{pre}$, $h=h_0$, whereas $h=h(1-\epsilon_{xx}^{pre}/2)$) otherwise. Tractions applied at the surface due to the line force generated by the liquid surface tension $\gamma_l$ take the form
\begin{equation}
	\begin{pmatrix}
    \sigma^h_{xz} \\
    \sigma^h_{zz} 
  \end{pmatrix} = \begin{pmatrix}
    \gamma_l\,\cos({\theta})\delta(x)\\
    \gamma_l\,\sin({\theta})\delta(x)
  \end{pmatrix} \Rightarrow 	\begin{pmatrix}
    \hat{\sigma}^h_{xz} \\
    \hat{\sigma}^h_{zz} 
  \end{pmatrix} = \begin{pmatrix}
    \gamma_l\,\cos({\theta})\\
    \gamma_l\,\sin({\theta})
  \end{pmatrix},
\end{equation}
where $\theta$ is the contact angle between the applied line force and the substrate's surface. Under the assumption of isotropic linear elasticity, \emph{e.g.}, $\sigma_{ij}=\lambda\epsilon_{kk}\delta{ij} + 2\mu\epsilon_{ij}$, tractions and displacements may be related via a coefficient matrix $S$ such that
\begin{align}
		\begin{pmatrix}
    \hat{\sigma}^h_{xz} \\
    \hat{\sigma}^h_{zz} 
  \end{pmatrix} &= \begin{pmatrix}
    S_{11} & S_{12} & S_{13} & S_{13} \\
     S_{21} & S_{22} & S_{23} & S_{23}
  \end{pmatrix} \hat{u}^h_i \nonumber \\
&= \begin{pmatrix}
    0 & \frac{i E}{2(1+\nu)}k & \frac{E}{2(1+\nu)} & 0 \\
    -\frac{i E\nu}{(1+\nu)(2\nu-1)}k & 0 & 0 & \frac{E(\nu-1)}{(1+\nu)(2\nu-1)}
  \end{pmatrix} \begin{pmatrix}
    B_2\,B_2^{-1} \\
    B_4\,B_2^{-1}
  \end{pmatrix} \begin{pmatrix}
    \hat{u}^h_x \\
    \hat{u}^h_z 
  \end{pmatrix}.
\label{eq:spring_const}
\end{align}
In shorthand notation, we write Equation~\ref{eq:spring_const} as
\begin{equation}
 \hat{\tau}^h_{i} = \hat{Q}_{ij}\hat{u}^h_j,
\label{eq:spring_const_short}
\end{equation}
and refer to $\hat{Q}_{ij}$ as a generic matrix of spring constants relating tractions $\hat{\tau}^h_{i}$ applied at the substrate's surface to surface displacements.

\subsection{Surface stresses}
\label{sec:Surf_Stresses}

We now proceed to furthermore take into account surface stresses which are counteracting tractions applied at the substrate's surface following \citep{Jerison:2011,Style:2012}. Starting from a force balance at the surface, we have
\begin{equation}
	[\hat{\sigma}_{ij}\hat{n}_j]^+_- = -(\delta_{ij}-\hat{n}_i\hat{n}_j)\hat{\Upsilon}_{ik,j},
\label{eq:force_balance_surf}
\end{equation}
where $\hat{\sigma}_{ij}$ denotes the Cauchy stress tensor within the bulk material, $\hat{\Upsilon}_{ik}$ represents the two-dimensional surface stress tensor and $\hat{n}_i$ is the surface normal in the undeformed configuration. For an isotropic linear elastic surface stress tensor, $\Upsilon_{ik}=\Upsilon^{0}\delta_{ik}+\lambda^s\epsilon_{mm}\delta_{ik}+2\mu^s\epsilon_{ik}$, Equation~\ref{eq:force_balance_surf} gives
\begin{align}
	[\hat{\sigma}^h_{xz}]^+_- &= [\hat{t_i}\hat{\sigma}_{ij}\hat{n}_j]^+_- \nonumber \\
&= -(\delta_{ij}-\hat{n}_i\hat{n}_j)(\hat{\Upsilon}_{ik}\hat{t}_k)_{,j} + (\delta_{ij}-\hat{n}_i\hat{n}_j)\hat{\Upsilon}_{ik}\hat{t}_{k,j} \nonumber \\
&= -(\delta_{ij}-\hat{n}_i\hat{n}_j)(\hat{\Upsilon}_{ik}\hat{t}_k)_{,j} \nonumber \\
&= k^2(\lambda^s+2\mu^s)\hat{u}^h_x
\label{Eq:Sigma_xz}
\end{align}
to first order in $\hat{u}_i$, with wave number $k$ \citep{Style:2018}.

To derive surface stresses counteracting normal tractions $[\hat{\sigma}^h_{zz}]^+_-$ in an equivalent way, we first note that
\begin{align}
	[\hat{\sigma}^h_{zz}]^+_- &= [\hat{n_i}\hat{\sigma}_{ij}\hat{n}_j]^+_- \nonumber \\
&= -(\delta_{ij}-\hat{n}_i\hat{n}_j)(\hat{\Upsilon}_{ik}\hat{n}_k)_{,j} + (\delta_{ij}-\hat{n}_i\hat{n}_j)\hat{\Upsilon}_{ik}\hat{n}_{k,j} \nonumber \\
&= 0.
\label{Eq:Other_sigma_zz}
\end{align}
We therefore derive an expression for the surface normal $\hat{\eta}_i$ in a perturbed configuration as
\begin{align}
  \hat{\eta}_i & = \begin{pmatrix}
    	-\hat{u}_{z,x} \\
    	-\hat{u}_{z,y} \\
	1 
  \end{pmatrix} + \mathcal{O}(u^2).
\end{align}
Using this relation, surface stresses counteracting tractions $[\hat{\sigma}^h_{zz}]^+_-$ may be derived as
\begin{align}
	[\hat{\sigma}^h_{zz}]^+_- &= [\hat{\eta_i}\hat{\sigma}_{ij}\hat{\eta}_j]^+_- \nonumber \\
&=  (\delta_{ij}-\hat{\eta}_i\hat{\eta}_j)\hat{\Upsilon}_{ik}\hat{\eta}_{k,j} \nonumber \\
&= k^2\Upsilon^0\hat{u}^h_z + \mathcal{O}(u^2).
\label{Eq:Sigma_zz}
\end{align}
Generalizing Equation~\ref{eq:spring_const_short} to the additional action of surface stresses therefore results in
\begin{equation}
 \hat{\tau}^h_{i} = (\hat{Q}_{ij}+\hat{T}_{ij})\hat{u}^h_j, \quad \text{with} \quad \hat{T}_{ij}=\begin{pmatrix}
    k^2(\lambda^s+2\mu^s) & 0 \\
    0 & k^2\Upsilon^0
  \end{pmatrix}.
\label{eq:spring_const_surf}
\end{equation}

\subsection{Anisotropic surface stresses based on pre-strain}

For an isotropic substrate subjected to uniaxial tension in $x$ and free boundaries in $y$ and $z$, strains are related as
\begin{equation}
	\epsilon^{pre}_{yy} = -\nu\epsilon^{pre}_{xx},
\end{equation}
with $\nu$ being the bulk's Poisson ratio. Resultant surface stresses follow as
\begin{align}
	\Upsilon^{pre}_{xx} &= (\lambda^s+2\mu^s-\nu\lambda^s)\epsilon^{pre}_{xx} \quad \text{and} \nonumber \\
	\Upsilon^{pre}_{yy} &= (\lambda^s-\nu2\mu^s-\nu\lambda^s)\epsilon^{pre}_{xx},
\end{align}
such that total surface stresses are given by $\Upsilon_{ik}=\Upsilon^{0}\delta_{ik}+\Upsilon^{pre}_{ik}+\lambda^s\epsilon_{mm}\delta_{ik}+2\mu^s\epsilon_{ik}$. To derive the fully three-dimensional profile of the wetting ridge, we introduce an arbitrary in-plane rotation $\psi$ connecting reference $(x,y)$ and rotated $(x',y')$ frames. Surface normals in the perturbed configuration are related as
\begin{equation}
	\hat{\eta}_i = R(\psi)_{ij^{'}}\hat{\eta}_{i^{'}} = \begin{pmatrix}
    -\cos{(\psi)}\hat{u}_{z^{'},x^{'}} \\
     -\sin{(\psi)}\hat{u}_{z^{'},x^{'}} \\
	1
  \end{pmatrix},
\end{equation}
whereby the corresponding gradient is given by
\begin{align}
	\hat{\eta}_{i,j} &= \hat{\eta}_{i,k^{'}}R(\psi)_{jk^{'}} \\
				    &= \begin{pmatrix} 
   	-\cos^2{(\psi)}u_{z^{'},x^{'}x^{'}} & -\cos{(\psi)}\sin{(\psi)}u_{z^{'},x^{'}x^{'}} & -\cos{(\psi)}u_{z^{'},x^{'}z^{'}} \\
	-\sin{(\psi)}\cos{(\psi)}u_{z^{'},x^{'}x^{'}} & -\sin^2{(\psi)}u_{z^{'},x^{'}x^{'}} & -\sin{(\psi)}u_{z^{'},x^{'}z^{'}} \\
	0 & 0 & 0 
  \end{pmatrix}. \nonumber
\end{align}

Extending Equation~\ref{eq:spring_const_surf} to take into account anisotropy in surface stresses based on an applied pre-strain hence results in
\begin{align}
 \hat{\tau}^h_{i} &= (\hat{Q}_{ij}+\hat{T}_{ij}+\hat{T}^{pre}_{ij})\hat{u}^h_j, \quad \text{with}  \nonumber \\
 \hat{T}^{pre}_{ij} &= \begin{pmatrix}
    0 & & & 0 \\
    0 & & & k^2(\Upsilon_{xx}^{pre}\cos^2{(\psi)}+\Upsilon_{yy}^{pre}\sin^2{(\psi))}
  \end{pmatrix}. 
\label{eq:spring_const_surf_aniso}
\end{align}

\subsection{Extension to finite deformation}

In the following, we extend the model to account for finite pre-strains of the surface and subsequently linearize about the pre-strained configuration preceding droplet deposition. We choose a two-dimensional compressible Neo-Hookean strain energy density 
\begin{equation}
	W^{NH} = \frac{\mu^s}{2}\left(\frac{I_1}{\text{det}(\mathbf{F})}-2\right) + \frac{\kappa^s}{2}(\text{det}(\mathbf{F})-1)^2,
\end{equation}
where $I_1=\text{tr}(\mathbf{F}^T\mathbf{F})$ denotes the first tensor invariant \citep{Holzapfel:2000}. Linearization about a given finite deformation gives the tangent matrix in spatial form as
\begin{equation}
	C_{ijkl} = \text{det}(\mathbf{F})^{-1}F_{jJ}F_{lL}C_{iJkL},
\end{equation}
with $C_{iJkL}=\partial(\partial W^{NH}/\partial F_{iJ})/\partial F_{kL}$ being the tangent stiffness matrix. Note the dependence on deformation gradient $\mathbf{F}$, which in turn renders the resulting  tangent matrix anisotropic in the present case of a given uniaxial pre-strain $\mathbf{F}=((1+\epsilon_{xx}^{pre},0),(0,1-\nu^s\epsilon_{xx}^{pre}))$. Resultant surface stresses upon subsequent droplet deposition follow as previously derived in Equations \eqref{Eq:Sigma_xz} and \eqref{Eq:Sigma_zz}, whereby now $\hat{\Upsilon}_{ij}=C_{ijkl}\epsilon_{kl}$.

\subsection{Calculated Surface Profiles}

\begin{figure}[t]
\begin{center}
    \includegraphics[width=1.\linewidth]{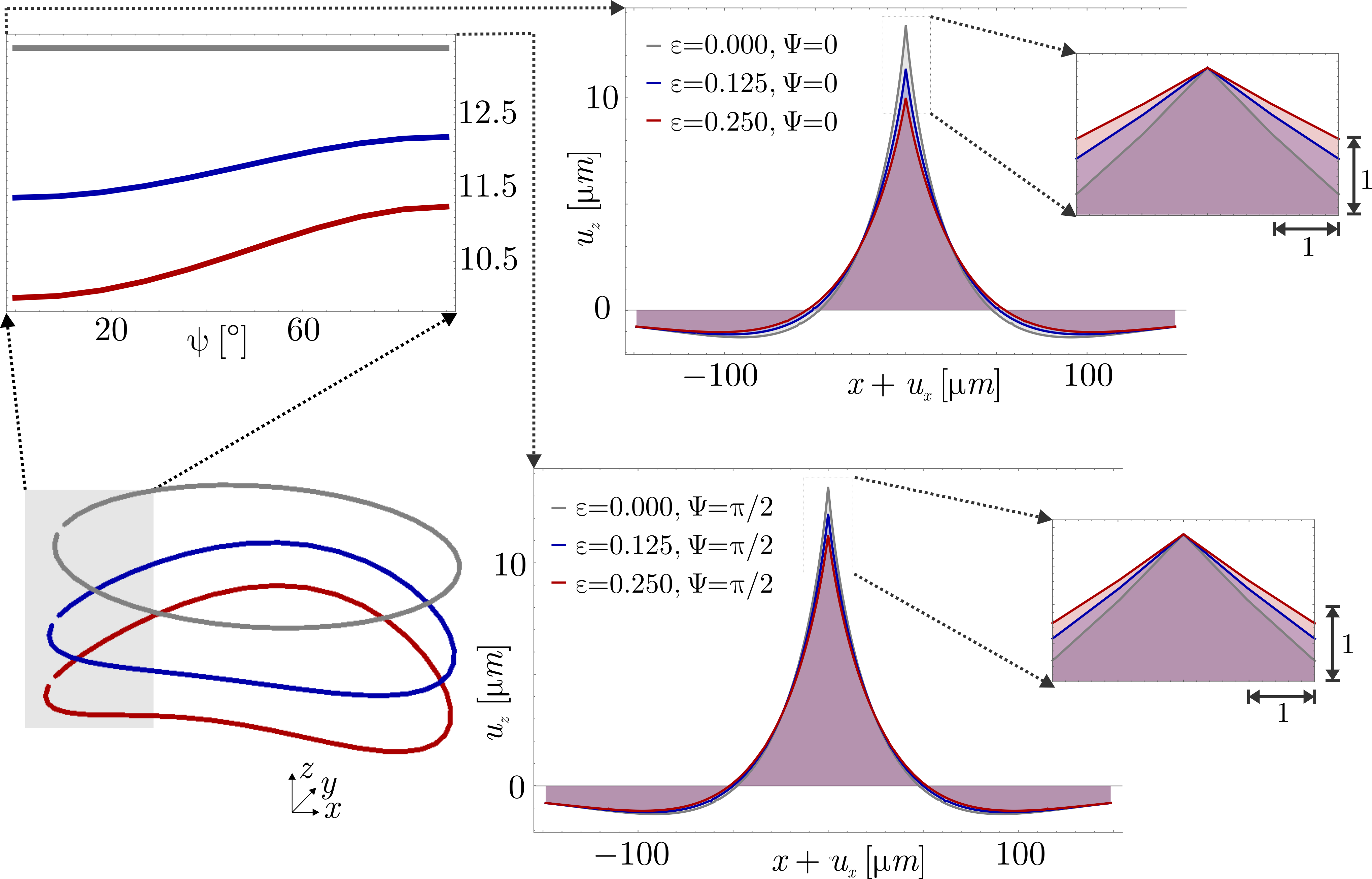}
    \caption{Static wetting on substrates exhibiting anisotropy in surface stresses based on applied pre-strains (linearized kinematics framework). Left: Wetting ridges along in-plane angle $\Psi$ under varying pre-strains of $\epsilon_{xx}=0.0$, $\epsilon_{xx}=0.125$ and $\epsilon_{xx}=0.25$. Right: Corresponding wetting profiles at $\Psi=0$ and $\Psi=\pi/2$ with inset illustrating the increase in opening angles with increasing pre-strain. Simulations employ values of $\Upsilon^0=0.02$\,N/m, $\mu^s=0.012$\,N/m, $\lambda^s=0.055$\,N/m and $E=2.88$\,kPa.
    }
\label{Fig:Ridge}
\end{center}
\end{figure}

Figure \ref{Fig:Ridge} illustrates theoretically predicted wetting profiles within the linearized kinematics framework at different strain levels. The wetting ridge decreases in height with increasing pre-strain levels due to a stiffer material response of the surface. 
This stiffening effect is more pronounced within the direction of applied pre-strain at $\Psi=0$. Similarly, opening angles of the wetting ridge increase with increasing pre-strain, showing a more distinct increase at $\Psi=0$. 

\section{Comparison with Experiments}

We now compare the the results of these calculations to experiments of Xu \emph{et al}.
To minimize systematic errors due to variations in material composition between different samples, we use an experimental subset of the data first published in \cite{Xu:2018} spanning one sample over a range of  pre-strain levels at positions $A,B$ at $\Psi=0,\pi/2$, respectively (see Figure~\ref{Fig:Droplet}).

\begin{figure}[t]
\begin{center}
    \includegraphics[width=1.\linewidth]{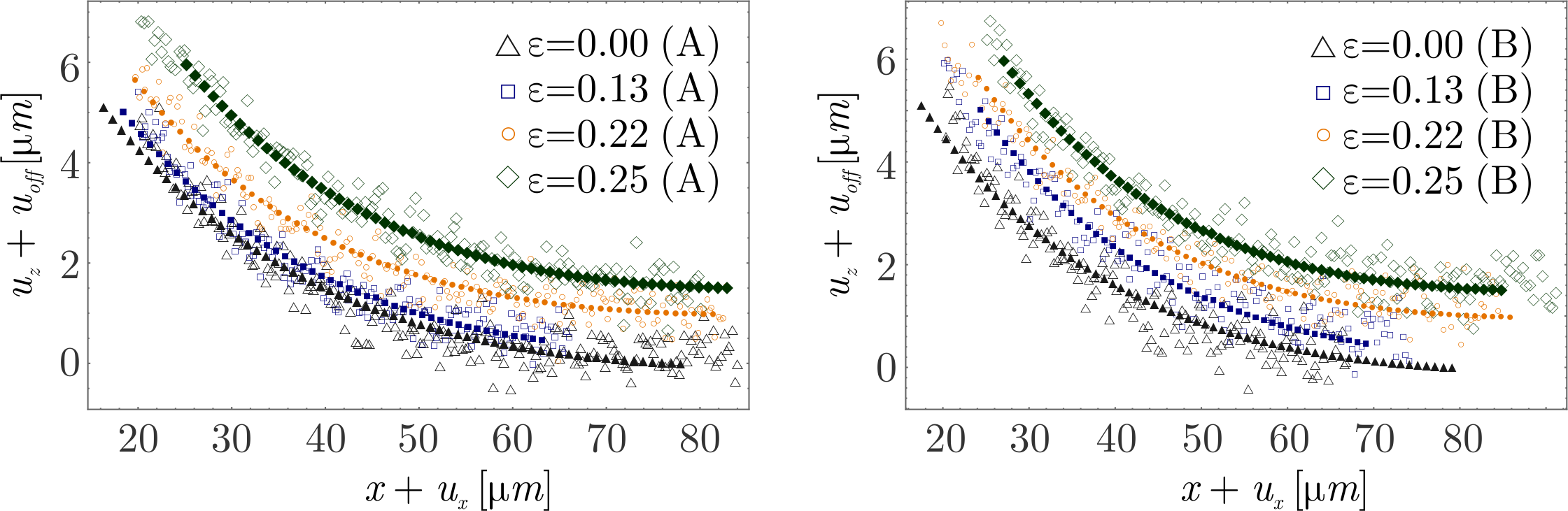}
    \caption{Profiles of substrate profiles far from the contact line with corresponding their corresponding best fits to the theory described here, with $E=2.88$\,kPa. Individual curves are shifted by an offset $u_{\text{off}}$ for better visibility.}
\label{Fig:Long}
\end{center}
\end{figure}

First, we fit the deformation of the substrate far from the contact line to determine the substrate elastic modulus.
In this region,  where the curvature is small, the liquid surface tension, $\gamma^l=0.041~\mathrm{N/m}$ \citep{Xu:2017}, is balanced by bulk elastic forces with contribution from solid capillary forces. 
Fitting ranges are chosen from $20$\,$\mu m$ to the final data point in each experimental data set. 
We find that far-field surface profiles for $4$ values of the pre-strain at both locations A and B can be simultaneously best-fit with a Young's modulus of $E=2.88$\,kPa. Figure~\ref{Fig:Long} depicts the tail behavior of experimental static wetting profiles and the corresponding fittings.  

\begin{figure}[t]
    \centering
    \begin{minipage}{.5\textwidth}
        \centering
        \includegraphics[width=0.99\textwidth]{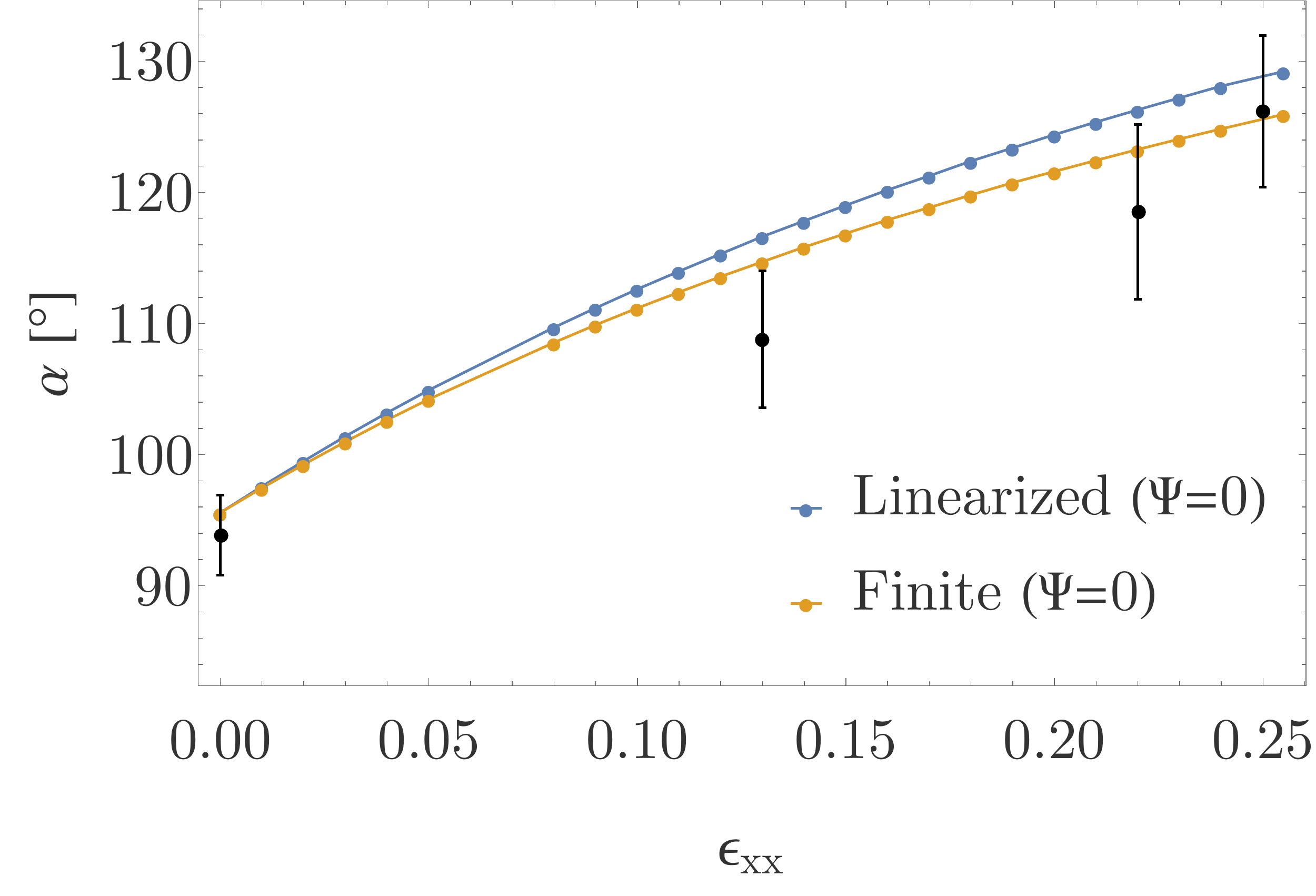}
    \end{minipage}%
    \begin{minipage}{0.5\textwidth}
        \centering
        \includegraphics[width=0.99\textwidth]{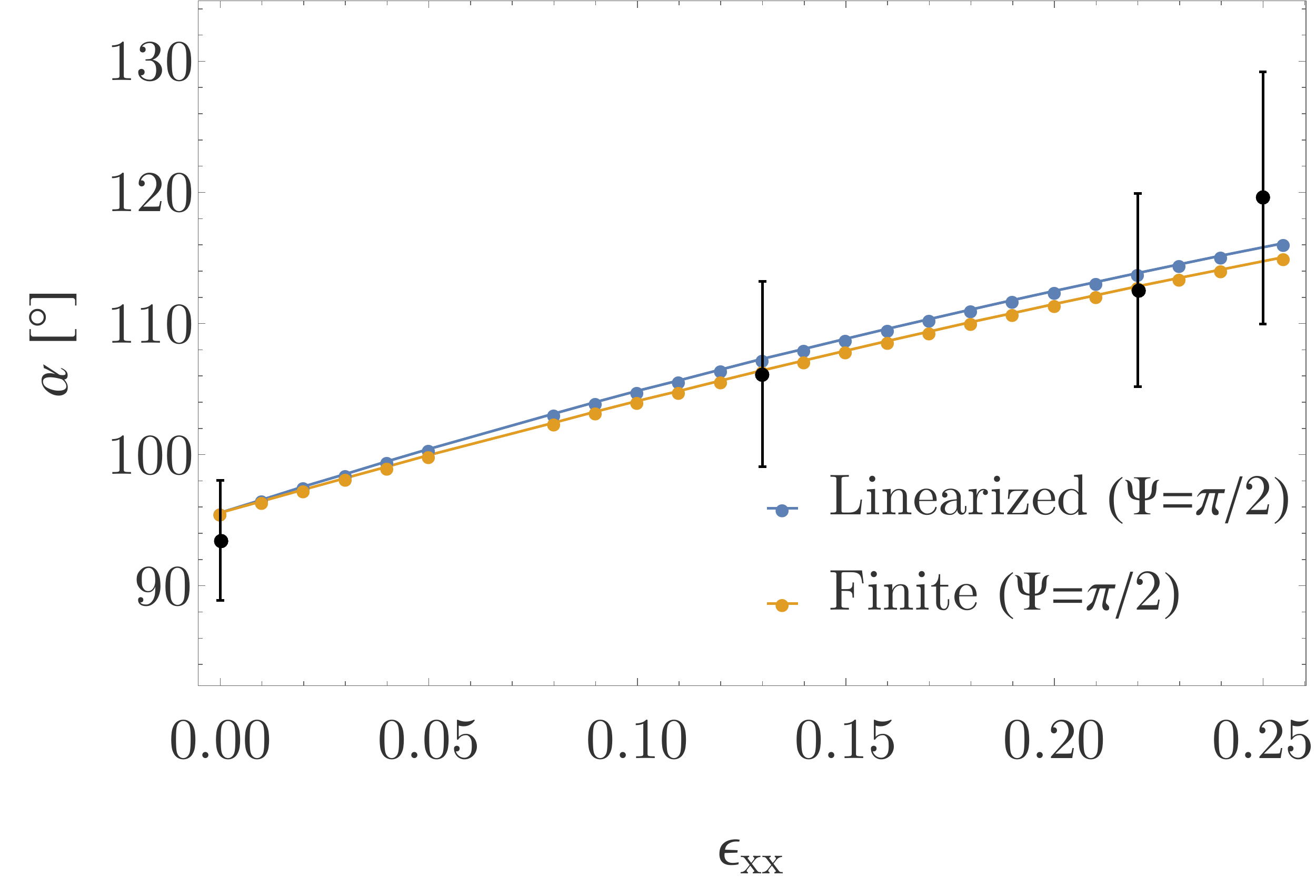}
    \end{minipage}
    \caption{Opening angles of the wetting ridge as predicted from linearized and finite kinematics theory in comparison to opening angles measured in static wetting experiments. Values are depicted for different pre-strain levels $\epsilon$ and compared both within the direction of applied pre-strain at $\Psi=0$ (A) as well as perpendicular to it at $\Psi=\pi/2$ (B).}
\label{Fig:Angles}
\end{figure}

Subsequently, surface elastic moduli $\lambda^s$ and $\mu^s$ are determined from a comparison of experimental and analytical opening angles at the tip of the wetting ridge, as surface stresses dominate the material behavior below the elastocapillary length. 
Opening angles are calculated by fitting surface profiles within $10$\,$\mu$m of either side of the ridge tip to quadratic polynomials. 
The position $x_0$ of the tip of the wetting ridge is included as a free parameter during the fitting procedure to account for possible shifts in experimental profiles.
Fitted profiles are depicted in Figure~\ref{Fig:Profiles}, and their opening angles are depicted in Fig. \ref{Fig:Angles}. 
We note that all tip profiles are well-fit within the linear regime.

\begin{figure}[t]
\begin{center}
    \includegraphics[width=0.6\linewidth]{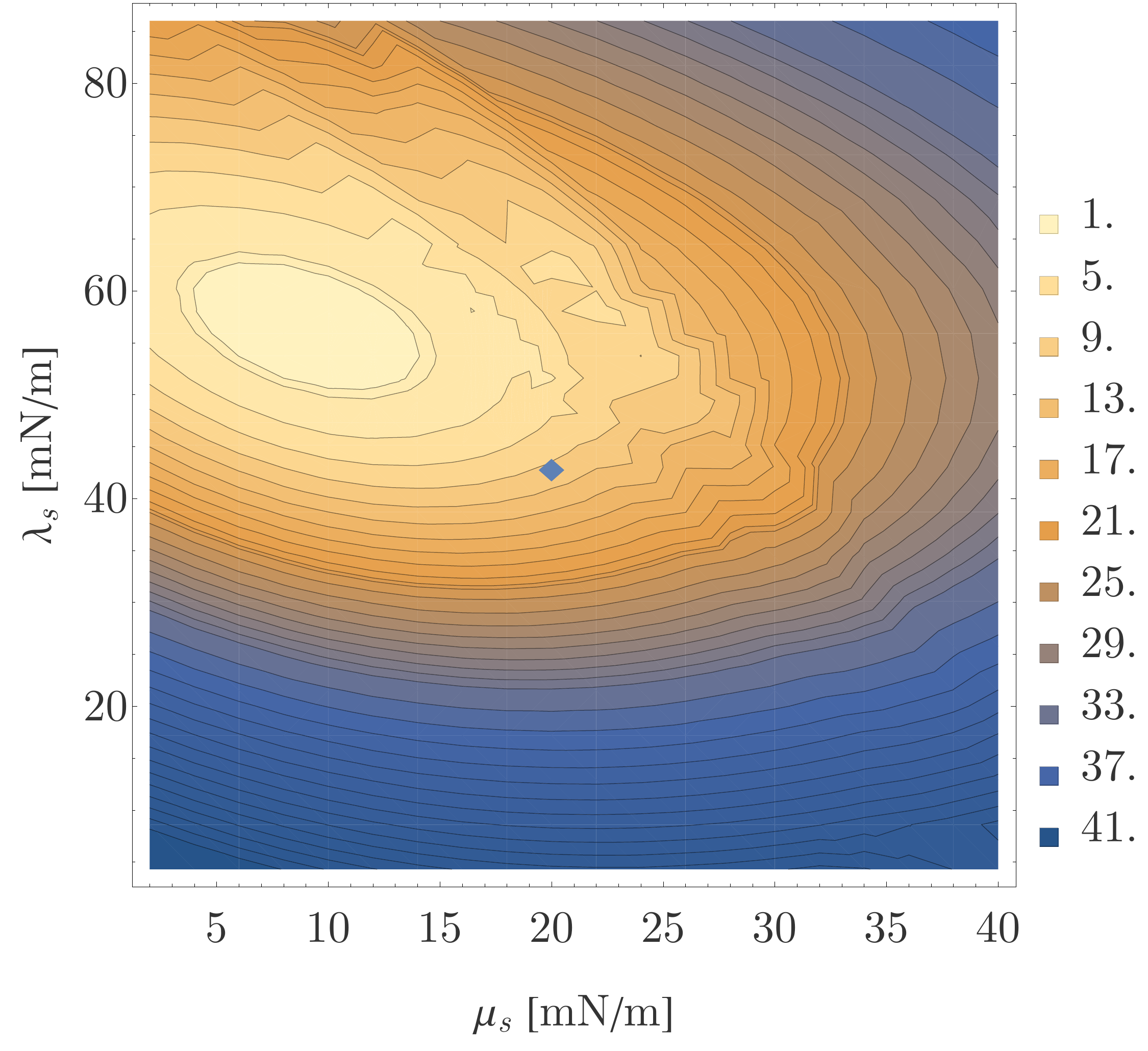}
    \caption{Contour plot showing reduced chi-square statistics $\chi^2$ applied to opening angles observed in experiments and simulations for varying surface elastic moduli. The marker located at $(\mu_s=20\,\text{mN/m},\lambda_s=43\,\text{mN/m})$ illustrates $\chi^2$ for surface elastic moduli published in \citep{Xu:2018}. 
    }
\label{Fig:Contours}
\end{center}
\end{figure}

To find best-fit values of surface elastic moduli, we seed the solution space with test points using a grid of varying discretization. 
Figure~\ref{Fig:Contours} illustrates reduced chi-square statistics $\chi^2$ applied to opening angles observed in experiments and simulations within the tested solution space. 
Lowest contour levels $\chi^2=0.95$ indicate that differences in opening angles obtained from experiments and simulations are in accordance with the experimental error variance. 
Resultant best-fit values are obtained as $\lambda^s=55\pm1$\,mN/m and $\mu^s=12\pm1$\,mN/m (with standard deviations corresponding to a $95\%$ confidence interval). 
Here, we fixed $\Upsilon^0=0.02$\,N/m and $\gamma^l=0.041$\,N/m as reported in previous investigations \cite{Xu:2017,Xu:2018}. 
This gives a surface Poisson's ratio, $\nu^s=\lambda^s/(2(\lambda^s+\mu^s))=0.41$.
In two dimensions, Poisson's ratio must lie on the interval  $\nu^s\in[0,1]$. 
While our specific values of $\lambda_s$ and $\mu_s$ differ from the values in \citep{Xu:2018}, our value of the biaxial stretching modulus $\Lambda=2(\lambda^s+\mu^s)$ is consistent from previously reported values \citep{Xu:2017,Xu:2018}.

A comparison of the experimental and best-fit opening angles are plotted in 
Figure \ref{Fig:Angles} as a function of applied strain.
As expected, both theories of linearized and finite kinematics predict the same opening angle at the tip of the wetting ridge for $\epsilon_{xx}=0$. In the direction perpendicular to the applied pre-strain, opening angles are recovered within experimental error measures in intermediate as well as high deformation regimes. Within the direction of applied pre-strain, however, linearized kinematics theory consistently overpredicts opening angles of the wetting ridge due to an inherent overstiffening of the material response. Using a finite kinematics model of surface elasticity softens the material response, resulting in decreased values of opening angles. This effect is more pronounced with increasing levels of applied pre-strain. Note that the chosen Neo-Hookean strain energy density allows for direct comparison to linearized kinematics theory, as both models of surface elasticity may be formulated in terms of model parameters $\mu^s$ and $\lambda^s$. Experimental opening angles have furthermore been compared to calculated values based on a two dimensional Yeoh strain energy density, for which results only display a marginal difference to the chosen Neo-Hookean model.

\begin{table}[t]
\begin{center}
  \begin{tabular}{ | l | c | c | c | r |}
    \hline
     $\epsilon_{xx}^{\text{pre}}$ & 0.0 & 0.13 & 0.22 & 0.25 \\ \hline
    \hline
    $\alpha^A_{\text{exp,sub}}$ & 94 $\pm$ 3 & 109 $\pm$ 5 & 119 $\pm$ 6 & 126 $\pm$ 6 \\ \hline
    $\alpha^A_{\text{theo,lin}}$ & 96 & 117 & 126 & 129 \\ \hline
    $\alpha^A_{\text{theo,fin}}$ & 96 & 115 & 123 & 126 \\
    \hline
    $\alpha^B_{\text{exp,sub}}$ & 93 $\pm$ 5 & 106 $\pm$ 7 & 113 $\pm$ 7 & 120 $\pm$ 9 \\ \hline
    $\alpha^B_{\text{theo,lin}}$ & 96 & 107 & 114 & 116  \\ \hline
    $\alpha^B_{\text{theo,fin}}$ & 96 & 106 & 113 & 115 \\ 
    \hline
  \end{tabular}
\caption{Opening angles $\alpha^{exp,sub}$ as determined from an experimental subset of data originally published in \citep{Xu:2018} and theoretically determined values $\alpha_\text{theo,lin}$ and $\alpha_\text{theo,fin}$. 
}
\label{tab:angles}
\end{center}
\end{table}

For a detailed comparison, Table~\ref{tab:angles} lists values of both analytically determined opening angles within linearized and finite kinematics theory ($\alpha_{theo,lin}$ and $\alpha_{theo,fin}$, respectively) as well as previously reported experimental results $\alpha_{exp}$ \cite{Xu:2018}. In addition, Table~\ref{tab:angles} lists opening angles calculated from an experimental subset of the data first published in \cite{Xu:2018}, which is used in determining best-fit surface elastic moduli $\lambda^s$ and $\mu^s$.

\section{Conclusions}

We have described a theory of strain-dependent surface stresses in the framework of static wetting by extending the generalized Flamant-Cerruti problem. 
Analytical solutions provide  fully three-dimensional surface profiles near a contact line, with an arbitary orientation relative to the direction of applied stretch. 
We found  that finite kinematics more accurately capture experimental  opening angles, especially with increasing levels of applied pre-strain. 

To minimize systematic errors, we limited our analysis to a subset of experiments from \citep{Xu:2018}, where a series of surface profiles are measured on a single substrate as a function of strain and location. 
While we recover re-analysis of these data returns the same value of the surface bi-axial stretching modulus, we find different values of the surface Lam\'{e} coefficients. 

The origins of surface elasticity in gels are unknown.
One simple model described in \cite{Style:2017} suggests that surface elasticity of gels can arise due to a  thin layer of material near the surface with a modulus much higher than the bulk.

In this picture, 2D Lam\'{e} moduli  (denoted as $s$) are related to the 3D Lam\'{e} of the putative stiff layer  (denoted as $l$) through $(\mu^s,\lambda^s)=(\mu^lh,\frac{2\mu^l\lambda^l}{\lambda^l+2\mu^l}h)$, where $h$  is the thickness of the stiff layer \cite{Benveniste:2007}. 
Substituting our measured values of $\lambda^s$ and $\mu^s$, we find $\frac{\mu^l}{\lambda^l}=-0.28$.
Since elastic moduli must be positive, this is an unphysical result.
This suggests  that the surface elastic moduli do not come from such a simple layer, or that nonlinear effects at the contact line \cite{Masurel:2019} need to be explicitly accounted for.

Possible extensions of the theory presented in this work include a re-formulation of the problem to allow for bulk elasticity within finite deformations. 
This extension may leverage further insights into the strain-dependence of surface stresses within the regime of large pre-deformations of the elastic substrate. 
An examination of different constitutive models of surface elasticity may furthermore give insights into the specific characteristics of strain dependence. 
As a secondary measure of comparison, theoretical results could be compared to local strains at the contact line. 
Current discrepancies between the predicted and measured in-plane displacements are based on the influence of the Laplace pressure difference of the droplet, and could hence be resolved by extending the theory to droplets of arbitrary size. 
Future experiments on a wider range of materials with different interfacial properties are furthermore needed to differentiate between the two hypotheses of strain-dependent surface stresses versus the effects of bulk nonlinearities being propagated to surfaces of constant surface energy \cite{Masurel:2019}. 
Possible experimental set-ups in future investigations could include a droplet deposited on a stretched soft solid substrate immersed in a second liquid phase. 
Varying second liquid phases would thus allow testing of a wide range of surface tensions at the contact line.

\section{Appendix}

The full form of components $B_1$ through $B_4$ of the matrix exponential $B_{ij}$ in equation~\eqref{eq:Bij} is given as
\begin{equation}
     B_{1} = \begin{pmatrix}
    \text{cosh}(kz) + \frac{kz\,\text{sinh}(kz)}{(2-4\nu)} & \frac{i(kz\,\text{cosh}(kz)-\text{sinh}(kz))}{4(-1+\nu)} \nonumber \\
    \frac{2i(kz\,\text{cosh}(kz)-\text{sinh}(kz))}{8\nu-4} & \text{cosh}(kz)+\frac{kz\,\text{sinh}(kz)}{(4\nu-4)} \nonumber \\
  \end{pmatrix} 
\end{equation}
\begin{equation}
     B_{2} = \begin{pmatrix}
     \frac{kz\,\text{cosh}(kz)+3\,\text{sinh}(kz)-4\nu\,\text{sinh}(kz)}{4k-4k\nu} & \frac{2iz\,\text{sinh}(kz)}{8\nu-4} \nonumber \\ \frac{iz\,\text{sinh}(kz)}{4(\nu-1)} & \frac{kz\,\text{cosh}(kz)-3\,\text{sinh}(kz)+4\nu\,\text{sinh}(kz)}{4k\nu-2k} \nonumber \\
  \end{pmatrix} 
\end{equation}
\begin{equation}
     B_{3} = \begin{pmatrix}
     \frac{k(-kz\,\text{cosh}(kz)+(4\nu-3)\,\text{sinh}(kz))}{4\nu-2}
       & \frac{ik^2z\,\text{sinh}(kz)}{4(\nu-1)} \nonumber \\
       \frac{2ik^2z\,\text{sinh}(kz)}{8\nu-4} & \frac{k(kz\,\text{cosh}(kz)+(4\nu-3)\,\text{sinh}(kz))}{4(\nu-1)}
       \end{pmatrix}
\end{equation}
\begin{equation}
     B_{4} = \begin{pmatrix}
     \text{cosh}(kz) + \frac{kz\,\text{sinh}(kz)}{(4-4\nu)} & \frac{2i(kz\,\text{cosh}(kz)+\text{sinh}(kz))}{8\nu-4} \nonumber \\
     \frac{i(kz\,\text{cosh}(kz)+\text{sinh}(kz))}{4(-1+\nu)} & \text{cosh}(kz) + \frac{kz\,\text{sinh}(kz)}{(4\nu-2)}
       \end{pmatrix}
\end{equation}

\pagebreak

%

\end{document}